\documentclass[journal]{./src/vgtc}                
\ifpdf
  \pdfoutput=1\relax                   
  \usepackage{graphicx}                
  \DeclareGraphicsExtensions{.pdf,.png,.jpg,.jpeg} 
\else
  \usepackage{graphicx}                
  \DeclareGraphicsExtensions{.eps}     
\fi%

\graphicspath{{figures/}{pictures/}{images/}{./}{src/}} 

\usepackage{microtype}                 
\PassOptionsToPackage{warn}{textcomp}  
\usepackage{textcomp}                  
\usepackage{mathptmx}                  
\usepackage{times}                     
\usepackage{cite}                      
\usepackage{tabu}                      
\usepackage{booktabs}                  
\usepackage{bm}
\usepackage{paralist}
\usepackage{enumitem}
\usepackage{tikz,float}
\usepackage[breaklinks,colorlinks,bookmarks=false]{hyperref}  
\usepackage{tabularx}
\usepackage{xparse}

\NewDocumentCommand{\statcirc}{ O{#2} m }{%
    \begin{tikzpicture}
    \fill[#2] (0,0) circle (3pt); 
    \fill[#1] (0,0) -- (180:3pt) arc (180:0:3pt) -- cycle; 
    \end{tikzpicture}
}

\onlineid{1426}

\vgtccategory{Research}
\vgtcpapertype{Applications}

\title{CohortVA: A Visual Analytic System for Interactive Exploration of Cohorts based on Historical Data}

\newcommand{\rc}[1]{\textcolor{black}{#1}}
\newcommand{\rw}[1]{\textcolor{black}{#1}}

\usepackage{xcolor}
\newcommand{\etal}{ {\textit{et al.}}}
\newcommand{\eg}{{\textit{e.g.}}}
\newcommand{\ie}{{\textit{i.e.}}}

\definecolor{TimeRange}{HTML}{AAB0BE}
\definecolor{Location}{HTML}{7891AA}
\definecolor{Affiliation}{HTML}{AD7982}
\definecolor{Relationship}{HTML}{8AA79B}
\definecolor{Celebrity}{HTML}{CA9087}
\definecolor{Entity}{HTML}{DDB997}
\definecolor{Blue}{HTML}{00ACFC}
\definecolor{Pink}{HTML}{FF8B8E}
\definecolor{Purple}{HTML}{6F5DAB}
\newcommand{\feature}[2]{\includegraphics[height=\fontcharht\font`\B]{Picture/feature/#1.pdf}\textit{\textcolor{#1}{#2}}}

\def\viewA{Scope Specification}
\def\viewB{Cohort Identification}
\def\viewBa{Cohort Feature Selection}
\def\viewBb{Cohort Analysis Provenance}
\def\viewC{Cohort Exploration}
\def\viewCa{Cohort Overview}
\def\viewCb{Figure Feature Validation}
\def\viewCba{Figure Feature}
\def\viewCbb{Figure Label}
\def\viewCc{Figure Event Validation}
\def\viewCca{Figure History}
\def\viewCcb{Event Ranking}
\def\viewCcc{Relationship Matrix}
\def\viewCcd{Event Map}
\def\viewCce{Event Timeline}
\def\viewD{Figure Details}

\newcommand{\fig}[1]{\autoref{fig:teaser}#1}
\newcommand{\figsub}[2]{\autoref{fig:teaser}#1#2} 
\newcommand{\figpipe}[1]{\autoref{fig:datapipeline}#1}
\newcommand{\figevent}[1]{\autoref{fig:event_design}#1}

\author{Wei Zhang, Jason K. Wong, Xumeng Wang, Youcheng Gong, Rongchen Zhu, Kai Liu, Zihan Yan, Siwei Tan, \\Huamin Qu, Siming Chen, and Wei Chen}
\authorfooter{
\item
W. Zhang, Y. Gong, R. Zhu, K. Liu, Z. Yan, S. Tan, and W. Chen are with the State Key Lab of CAD\&CG, Zhejiang University. W. Chen is also with the Laboratory of Art and Archaeology Image (Zhejiang University), Ministry of Education, China. E-mail: \{zw\underline{~~}yixian, 21951121, zrcccrz, kai.lk, zihanyan, siweitan, chenvis\}@zju.edu.cn. Wei Chen is the corresponding author.

\item
J. Wong and H. Qu are with the Hong Kong University of Science and Technology. Email: \{kkwongar, huamin\}@cse.ust.hk.

\item
X. Wang is with TMCC, CS, Nankai University. E-mail: wangxumeng@nankai.edu.cn.

\item
S. Chen is with Fudan University. E-mail: simingchen@fudan.edu.cn.

}

\shortauthortitle{Biv \MakeLowercase{\textit{et al.}}: Global Illumination for Fun and Profit}

\abstract{
 In history research, cohort analysis seeks to identify social structures and figure mobilities by studying the group-based behavior of historical figures. Prior works mainly employ automatic data mining approaches, lacking effective visual explanation. In this paper, we present CohortVA, an interactive visual analytic approach that enables historians to incorporate expertise and insight into the iterative exploration process. The kernel of CohortVA is a novel identification model that generates candidate cohorts and constructs cohort features by means of pre-built knowledge graphs constructed from large-scale history databases. We propose a set of coordinated views to illustrate identified cohorts and features coupled with historical events and figure profiles. Two case studies and interviews with historians demonstrate that CohortVA can greatly enhance the capabilities of cohort identifications, figure authentications, and hypothesis generation.
}


\CCScatlist{ 
 \CCScat{K.6.1}{Management of Computing and Information Systems}%
{Project and People Management}{Life Cycle};
 \CCScat{K.7.m}{The Computing Profession}{Miscellaneous}{Ethics}
}

\teaser{
  \centering
  \includegraphics[width=\linewidth]{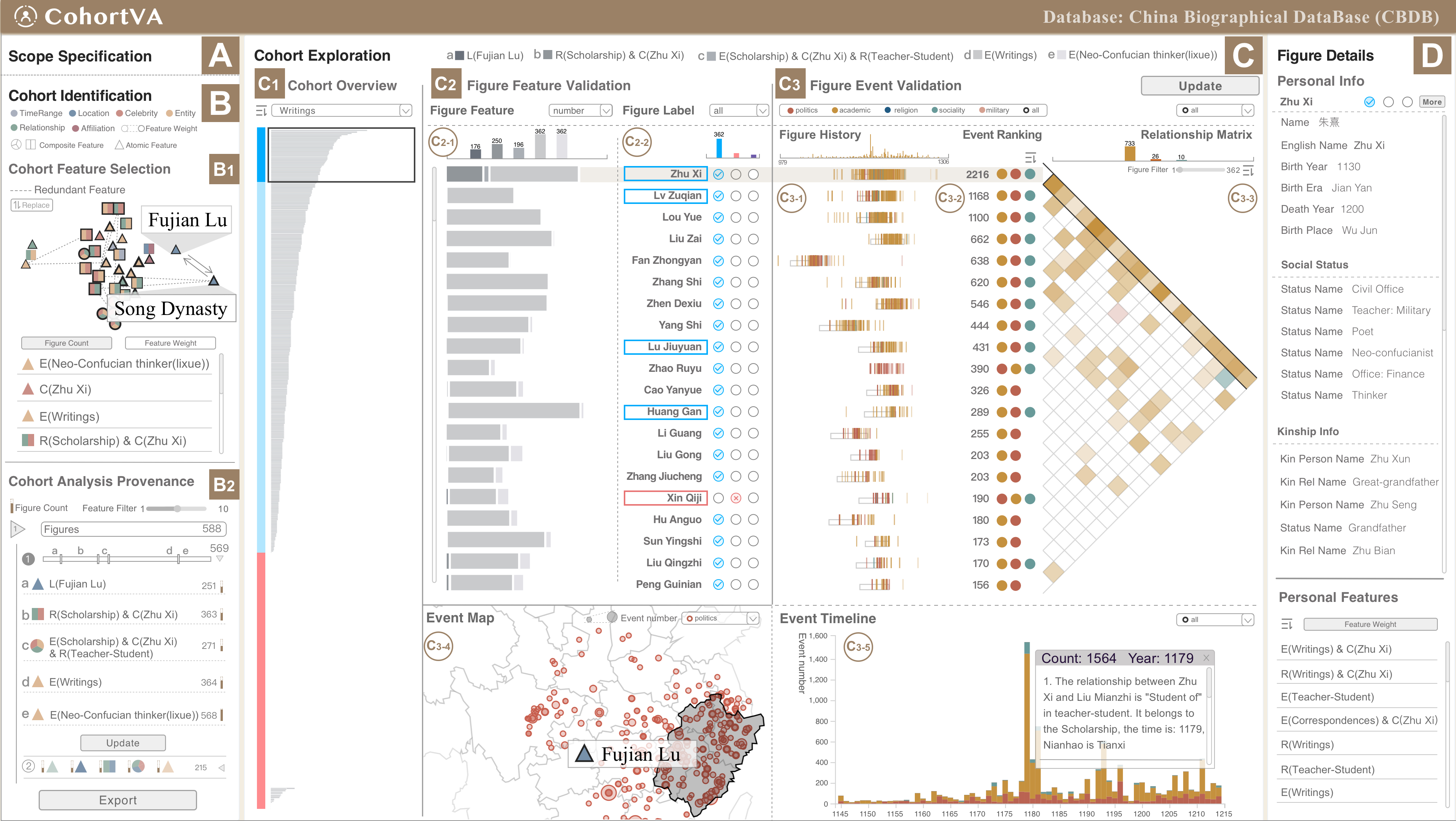}
  \caption{
  CohortVA streamlines the iterative cohort analysis workflow and provides visual interpretation for model results. (A) The \viewA~Component configures the initial research scope. (B) The \viewB~Component visually summarizes recommended cohort and feature candidates. (C) The \viewC~Component provides multi-faceted visualizations to validate the selected cohort from multiple perspectives. (D) The \viewD~Component shows the historical figure's profile.
  }
  \label{fig:teaser}
}

\vgtcinsertpkg

\begin{document}
\maketitle


\maketitle
\section{Introduction} 

Cohort analysis is a crucial research area in history studies \rw{known as \textit{prosopography}}, which can inspire the interpretation of the historical process. Here, a \textit{cohort} refers to a group of figures that engage in common activities or have frequent interactions~\cite{ryder1985cohort}. \rc{A prosopographical study typically focuses on one cohort and explores its \textit{concept},~\ie, the set of supplementary features that describes a cohort (\eg, political identities, relationship networks, and social structure)~\cite{bol2012gis, turchin2018quantitative}.}
\rc{For instance, based on cohort analysis, Beard~\cite{beard1913economic} revealed that the United States' founding fathers were closely tied with not only leading the American revolution together, but also personal financial interests. This new concept has provoked widespread discussions throughout the U.S. Federal Constitution from a financial perspective. More importantly, it has inspired subsequent works to adopt cohort analysis in suggesting novel concepts and interpretations of well-known cohorts~\cite{wilentz2007american,iggulden2008loyalist}.}

Recent advancements in digital humanities have greatly relieved historians from manual data collection and labeling. One remarkable progress is that large-scale historical databases, such as China Biographical DataBase~\cite{cbdb} and China Government Employee Database-Qing~\cite{chen_campbell_ren_lee_2020}, have been carefully built and widely used. Leveraging them in cohort analysis poses new challenges because the data sizes and varieties overwhelm analysts' capabilities~\cite{van2012if}. Automatic analysis approaches can alleviate these difficulties yet are far from ideal. First, existing tools focus partially on extracting and analyzing features, but merely support correlation analysis in the context of social structure. For example, Netdraw~\cite{borgatti2002netdraw} and Worldmap~\cite{guan2012worldmap} can only construct basic features from relation networks and geographic locations. Second, most tools are inefficient in integrating human intelligence and supporting iterative exploration. Third, the lack of interpretability prevents historians from effectively verifying obtained results. Historians need additional effort to authenticate results by referring to other documents.

As such, it is highly desirable to integrate domain knowledge and expert hypothesis into the exploration procedure within an intuitive visual interface. For that, we closely worked with historians to observe their behaviors and needs, and identified two challenges: 1) Historians must devote a significant amount of time to synthesizing and cross-checking findings of various figures from a large body of historical literature. It is very time-consuming because figure profiles and features cover various perspectives, such as native place, gender, and changes in official positions. 2) The entire process is repetitive because they may constantly shift the goals. Starting from a familiar targeted group, they might discover interesting patterns and recursively explore and study relevant figures. Considering their exploration process and such obstacles, this is a missed opportunity to provide a visualization-driven solution for efficient analysis and reasoning.

The kernel of our solution, CohortVA, is a cohort identification model that can automatically identify cohorts and their figures. Specifically, given an initial group of interested figures, our model constructs a knowledge graph of related documentary descriptions. It extracts and fuses the common features as the cohort's concept. Based on weakly-supervised learning, our model identifies new cohorts by proposing different cohort concepts as recommendations.
The recommended cohorts and their features are visually investigated within the integrated visual interface from multiple perspectives,~\eg, location, time, and relationship.
In particular, historians can study feature details, reason relations among figures, and progressively refine features and figures of the targeted cohort. 
Case studies and expert interviews demonstrate that CohortVA frees historians from their heavy workloads and improves the analysis performance. This study makes the following contributions:
\begin{itemize}[noitemsep,topsep=0pt]
    \item{We propose a cohort identification model that utilizes weakly-supervised learning to free historians from manual annotation, querying, and cross-checking.}
    \item{We develop a visual analytic approach, enabling historians to study cohorts and concepts interactively.}
    \item{We conduct case studies and expert interviews to demonstrate the effectiveness and usefulness of our approach.}
\end{itemize}

\section{Related Work}

In this section, we review the relevant works in history-oriented visual analysis and visual analytics for cohort studies. 

\subsection{\rw{History-Oriented Visual Analysis}}
Recently, visual analysis has been widely applied to historical data~\cite{8617736}.
\rc{History-oriented approaches can be categorized as phenomenon-based and theory-based.
A \textit{phenomenon} refers to a recorded historical event, while a \textit{theory} explains why one or more phenomena occurred.}

\rc{\textbf{Phenomenon-based} research analyzes the phenomena of historical entities such as figures~\cite{JanickeFS16,9695348}, events~\cite{7192676,9447222}, and cultures~\cite{benito2017exploring,feng2022ipoet, castermans2017glottovis}. They focus on a few instances in great detail to find hidden patterns and correlations for the phenomenon. Zhang\etal~\cite{ZhangMPC21} contextualized poems from the Chinese Song dynasty with the poets' life stories. The Svoboda Diaries Project~\cite{chen2019grounding} uses a person's diaries to recreate the personal experience in Ottoman Iraq. 
However, they fail to provide a holistic overview for interpreting and generalizing similar phenomena.}

\rc{\textbf{Theory-based} research deduces coherent explanations from several phenomena.} To explore the similarities among social structures, Turchin\etal~\cite{turchin2018quantitative} collected and summarized the characteristics (\eg, social scale, economy, and information systems) of 414 societies from 30 regions. \rw{Regarding a smaller social unit, GeneaQuilts~\cite{bezerianos2010geneaquilts} presents large family trees in an interactive diagonal matrix to study genealogical relationships. Similarly, GenealogyVis~\cite{7909028} explores family structure via correlations between family development and social environment.} \rc{These works demonstrate the benefits of analyzing social structure from a group perspective. Another popular domain investigates the social mobility aggregated by individual movements. For example, Bol~\cite{bol2012gis} adopted geospatial analysis to study how the Southern Chinese intellectual-social movements spread through the $12^{th}$ century. Khulusi\etal~\cite{8781573} used network analysis and a novel visualization design to define groups interactively for musicians' biography. CareerLens~\cite{9382844} and ACSeeker~\cite{9552870} utilize time-series analysis to explore career trajectories.} 

\rc{In this work, we mainly follow the theory-based approach to assist historians in developing the concepts behind cohorts. We mine features from large-scale multi-dimensional data and fuse them to propose concept candidates. These concepts characterize the cohort and form the basis for understanding the phenomena. We also borrow ideas from phenomenon-based approaches to let users verify the cohorts and contextualize the concepts with detailed historical events.}

\subsection{\rc{Visual Analytics for Cohort Studies}}

Cohort studies are widely used across various domains, such as medicine~\cite{zhang2015iterative,raidou2018bladder} and biology~\cite{8440823}. For instance, \rw{CoCo~\cite{malik2015cohort} integrates statistical and visual analysis for the medical experts to classify and compare cohorts' time series. PhenoStacks~\cite{7534774} simplifies ontological topologies and explores symptom similarities among inter-groups and intra-groups of patients}. 
\rc{However, these works focus on analyzing and classifying multiple entities into cohorts with existing and clear definitions. They cannot be directly adopted to historical cohort analysis, which emphasizes the identification of new and vaguely defined cohorts. In addition, they cannot iteratively refine the cohort concepts, which usually only become more evident during the exploratory analysis.}

The development of historical databases has contributed significantly to prosopography~\cite{jeffreys2006prosopography,bol2004china}. \rw{Historians widely adopt data visualization tools to explore cohort characteristics from digital records. For example, Gephi~\cite{grandjean2015gephi}, Netdraw~\cite{borgatti2002netdraw}, and Worldmap~\cite{guan2012worldmap} leverage basic visualization, such as force-directed graphs and choropleth maps, to help historians organize data intuitively.}
\rc{However, these tools fail to integrate multi-dimensional data and features effectively for cohort analysis.
Historians need to spend much time cross-validating discoveries about cohorts in the extensive historical literature.}

\rw{Besides visualization, machine learning techniques have significantly empowered cohort visual analysis in terms of efficiency~\cite{devi2016analysis}}. For example, Zhao\etal~\cite{zhao2011new} used non-binary hierarchical trees and overlapping clustering to shortlist important clusters, reducing the cost of manual selection. \rc{However, historians are often confused about the semantic meaning behind the automatic outputs,~\ie, the extracted latent features. For instance, Franke\etal~\cite{franke2019confidence} pointed out that confidence was the first-class attribute of historians for data adoption. They have thus raised awareness of improving interpretability for wider tool adoption.}

\rc{The closest work to ours is \textit{PK-clustering}~\cite{pister2020integrating}. It captures the user's prior knowledge as a set of incomplete clusters, then runs multiple clustering algorithms and visually compares the ensemble results for the user's decisions. It enhances interpretability by keeping human-in-the-loop for each analytical iteration.
Although PK-clustering and CohortVA both rely on users' interactions with the interim results, there are still a few key differences compared with our work. 
First, PK-clustering performs typical clustering tasks on medium-sized datasets with 50-500 entities. 
In contrast, CohortVA identifies a cluster as one cohort from a large-scale relational database having over $500$K historical figures. 
We proposed a weakly-supervised learning model and interlinked views to address the difference in scale.
Second, instead of visualizing the results alone, we provide interpretable features that corroborate the cohort and auxiliary domain information to explain the results. These quickly validate the cohort composition.}

\section{Background}
\label{Background}

In this section, we introduce the domain background information and outline the requirements obtained through interviewing domain experts.
\rw{To characterize domain problems and formulate system requirements, we have worked closely with five experienced historians over the past two years. Three of them~(H1-H3) are in charge of the China Biographical DataBase~(CBDB)~\cite{cbdb}. H1 and H2 are committee members, and H3 is a professor leading cohort analysis research projects based on CBDB. The other two~(H4 and H5) are Ph.D. students who use CBDB to study the mobility of historical cohorts and the history of the Chinese Song dynasty, respectively. Our collaboration consists of five phases: data acquisition~(\autoref{Data}), task analysis~(\autoref{Task}), model design~(\autoref{sec:model}), system design~(\autoref{Visual}), and system evaluation~(\autoref{Evaluation}).}%

\subsection{Data Description}
\label{Data}

This study employs CBDB, a large-scale open-sourced relational database, to study cohorts in Chinese history. CBDB contains enriched biographical records for over $500$K historical figures, spanning from the 7\textsuperscript{th} century to the 19\textsuperscript{th} century. The records have been entered and validated by experienced historians. They are multidimensional, covering four main information types: 

\begin{itemize}[itemsep=2pt,topsep=0pt,parsep=0pt,leftmargin=*]
    \item{\emph{Figures' attributes:} the basic personal information of figures, including birth year, death year, \rw{place of birth, place of burial,} gender, offices, ethnicity, writing, etc.}
    \item{\emph{Figures' relationships:} the social information among figures, including domestic, political, social, and academic relationships, such as colleagues, friendships, kinship, peers, teacher-student, etc.}
    \item{\emph{Historical events:} the event description with the time, place, and figure information. \rc{As in \autoref{fig:datapipeline}A2, Zhang Jiuling served as a prefectural aide in 737 for the Jingzhou prefecture, Shannan circuit.}}
    \item{\emph{Supplementary information:} the detailed description of personal attributes, such as location coordinates, time duration of different dynasties, background knowledge of each dynasty (the governor information and the bureaucratic hierarchy), etc. The amount of information for each figure varies between 5 and 500 records.}
\end{itemize}

\begin{table*}[tb]
    \renewcommand\arraystretch{1.5}
	\centering  
	\caption{Explanation and extraction models for the six atomic features. Here, descriptions refer to those about the selected figures.}  %
	\label{tbl:atomic_feature}
	\begin{tabular}{p{2cm}|p{4cm}|p{11cm}}
		\hline
		Atomic feature & Explanation & Extraction approach\\
		\hline
        TimeRange   & The time period in which an event happened & Cluster the years occurred in the descriptions with DBSCAN~\cite{khan2014dbscan}. Then, select the year ranges with more than 30\% occurrences as \textit{TimeRange} features.\\
        Location    & The location where an event happened & Select the top-three locations in the descriptions as \textit{Location} features.\\
        Affiliation & The government institution where figures held positions & Select the top-three offices in the descriptions as \textit{Affiliation} features.\\
		Relationship & The relationship (\eg, teacher-student) associating two figures & Construct a strongly connected relationship graph based on the descriptions. Then, select the relationships in communities with over five members as \textit{Relationship} features.\\
        Celebrity   & The person mostly connected by others & Link any two figures if they appear in the same description. Then, select the figures linked with over 30\% of the selected figures as \textit{Celebrity} features.\\
        Entity   & The entity, like occasions and writing & Link a figure with an entity if they appear in at least one description. Then, select the entities linked with over 30\% of the selected figures as \textit{Entity} features. Only those entities not covered by other features are selected.\\
		\hline
	\end{tabular}
\vspace{-0.2in}
\end{table*}

\subsection{Requirement and Task Analysis}
\label{Prosopography}
\label{Task}

\begin{figure}[htbp]
\centering
\includegraphics[width=\linewidth]{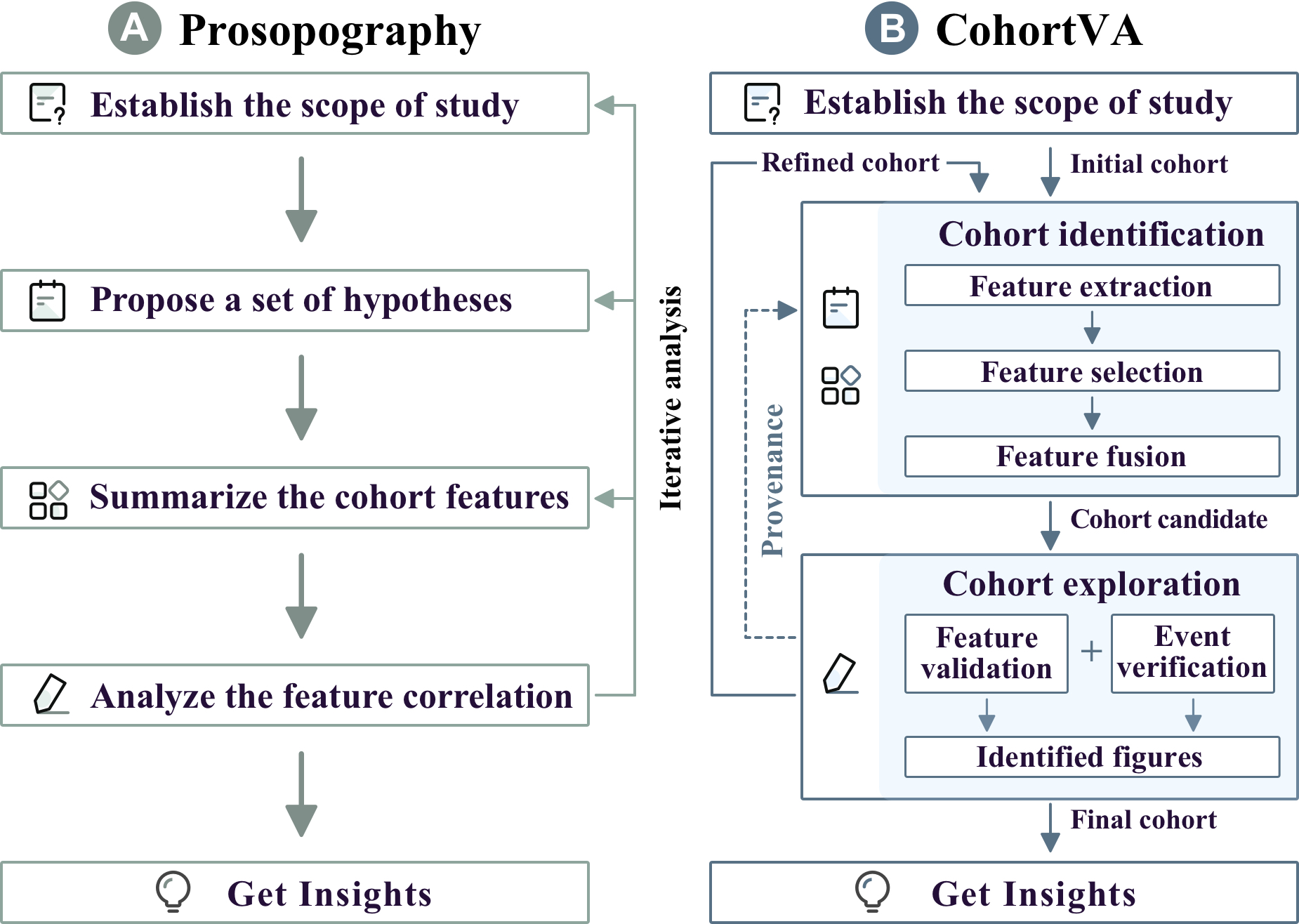}
\caption{A comparison between (A) traditional prosopography workflow and (B) proposed workflow applied in CohortVA.}
\label{fig:workflow_comparison}
\end{figure}

We interviewed our collaborators about their traditional workflow \rw{to guide our system design. 
The traditional historical cohort research follows a mature paradigm \cite{stone1971prosopography,bol2012gis}, including hypothesis formulation, feature summarization, and correlation analysis as shown in~\autoref{fig:workflow_comparison}A}.
Our collaborators indicated that the traditional analysis process is energy-exhausting and time-consuming. 
We summarised the system requirements and design tasks that empower historians with enhanced cohort analysis capabilities, geared toward their challenges.

{\bf Identifying cohorts from a large-scale historical database.} The first step of cohort analysis is to define the scope of the study and formulate corresponding hypotheses. To find the targeted scope, historians need to browse and filter the massive data materials. They then need to perform a heuristic search based on their experience and hypotheses. Verifying a hypothesis sometimes costs researchers a few months to read related materials repeatedly. The system should provide efficient cohort identification based on users' interests.
    \begin{enumerate}[label=\textbf{T{\arabic*}}, start={1}, nolistsep]
    \item {\bf Specify the initial research scope}. \rc{Historians' research interests are motivated by different perspectives. 
    For example, supporters of the great man theory would be interested in a few core figures, while those supporting geographical determinism tend to look for location correlations~\cite{bol2012gis}.
    Moreover, the differences in prior knowledge about cohorts lead to different specifications. We should support different research orientations by specifying the initial groups in multiple ways.}
    
    \item {\bf \rc{Generate automatic cohort identification results.}} 
    \rc{Traditional prosopography workflow in identifying groups~(\autoref{fig:workflow_comparison}A) includes browsing massive historical records within the research scope. Automatic cohort identification can alleviate the burden. Since historians might not know everything about a certain dynasty, the automatic method should also account for reasonable ambiguity. Moreover, historians appreciate various cohort candidates complementing cohort analysis from multiple angles.}
    \end{enumerate}
    
{\bf Providing visual interpretation for cohort identification results.} Historians need to cross-check the identified cohorts before using them for further research. The system should provide a set of visual interpretations to assist in result verification.
    \begin{enumerate}[label=\textbf{T{\arabic*}}, start={3}, nolistsep]

    \item {\bf \rc{Validate the concept and features that define a cohort}.}
    \rc{Historians seek variables of significance to explain a cohort's phenomena. 
    The identified cohort includes the list of figures and the cohort concept. 
    Historians need to validate whether the cohort concept adequately covers the associated figures. 
    Moreover, to find the most suitable concept, historians also want to examine alternative features of the identified cohort.
    }
    
    \item {\bf \rc{Verify the cohort from the organized historical event information.}} 
    \rc{H1-H5 emphasize that they always search for additional evidence to verify the results of automated methods. It is necessary to present historians with rich contexts that historians frequently reference, such as geographic locations and social network relationships. They can cross-check the identified cohorts from the detailed historical event information.}
    
    \item {\bf Inspect individual figures.} 
    \rc{Analyzing individual profiles help historians interpret the identified cohort at the most detailed level. All mined features and event evidence can be provided for inspiration. The system should display figure profiles from CBDB with spatial-temporal information and descriptions of social relationships. Directing users to the original sources outside the system should also enhance their trust level in the result.}
    \end{enumerate}
    
\textbf{Supporting iterative cohort analysis.}
Cohort analysis processes are naturally iterative, and the system should be able to support them.
    \begin{enumerate}[label=\textbf{T{\arabic*}}, start={6}, nolistsep]
    \item \rc{
    {\bf Adapt to the revising research interests.}
    Historians refine their research scopes iteratively during the analytical process. 
    Due to different research interests, they might disagree on how features are fused and cohorts are composed. 
    For example, Sima Guang and Su Shi are famous writers and political rivals in the Song dynasty.
    When historians decide to focus on one particular political party, cohorts containing both figures should be discarded, and corresponding features should be remodeled.
    Our system should update the cohort composition and concept recommendations accordingly, to reflect the revised research interests.
    }
    
    \item {\bf Track analytic provenance.}
    \rc{
    One of the most mundane tasks in prosopography is the inclusion of newly discovered variables. Upon new discoveries, historians need to trace back to previous steps to test what-if scenarios, accompanied by revisiting the same documents repeatedly. We should support revisiting previous progress to compare different cohort identification results.
    }
    \end{enumerate}

\section{Cohort Identification Model}
\label{sec:model}

The cohort identification model discovers potential cohorts from CBDB based on specified figures and features.
As illustrated in~\autoref{fig:datapipeline}A, the model contains four steps: \rc{1) generate the knowledge graph and descriptions}, 2) extract common features from the initial figures, 3) select the features by their significance, and 4) fuse the selected features as the cohort concept and filter figures by this concept.%

\subsection{\rc{Knowledge Graph and Description Generation}}
\rc{We preprocessed the raw historical data in CBDB by converting them into a knowledge graph~(\autoref{fig:datapipeline}A1), where the information is stored in nodes~(\ie, figures and entities) and edges~(\ie, relationships). 
In CBDB, figures and entities are structured as rows, and their relationships are as foreign keys. 
For instance, in the [POSTED\_TO\_OFFICE\_DATA] table, an [Posting] event is connected to [Zhang Jiuling] and an office position entity [Prefectural Aide].
The two triplets, (Zhang Jiuling, do, Posting) and (Posting, officeIs, Prefectural Aide), are inferred and inserted into the knowledge graph.
We identified 28 node types and 27 edge types, and built a knowledge graph with around 1M nodes and 5M edges.}

\rc{
We adopted the meta-path2vec~\cite{dong2017metapath2vec} algorithm to generate descriptions around the figure entities.
Collaborating with historians, we summarized 15 description templates that express interpretable and descriptive information, such as politics, occupation, and social relationships. 
Meta-path2vec leverages the templates and the sequential structure of nodes connected by edges in the knowledge graph. 
Compared with the conventional random walk method~\cite{lawler2010random}, it uniformly generates descriptions regardless of nodes' degrees, avoiding the probability-imbalance issue on different node types.
The generated descriptions represent a figure's semantic contexts (see \autoref{fig:datapipeline}A2 for an example).
Lastly, we obtained about 500K figures and 1M descriptions.
}

\subsection{Feature Extraction}
\label{sec:feat_extract}
\rw{We refer to the figure-related characteristics extracted from the descriptions as \textit{features}. 
Following the conventional cohort study paradigm~\cite{bol2012gis}, we categorized figures' characteristics in CBDB into six types of \textit{atomic features}, as shown in~\autoref{tbl:atomic_feature}. To represent more complicated contexts, a \textit{composite feature} is generated from multiple atomic features via the \textit{and} logical combination. For instance, the composite feature $[Location(Jingzhou)~\&~TimeRange(737)]$ indicates that the figures visited the Jingzhou prefecture in 737.}

\begin{figure*}[ht]
\centering
\includegraphics[width=1.0\linewidth]{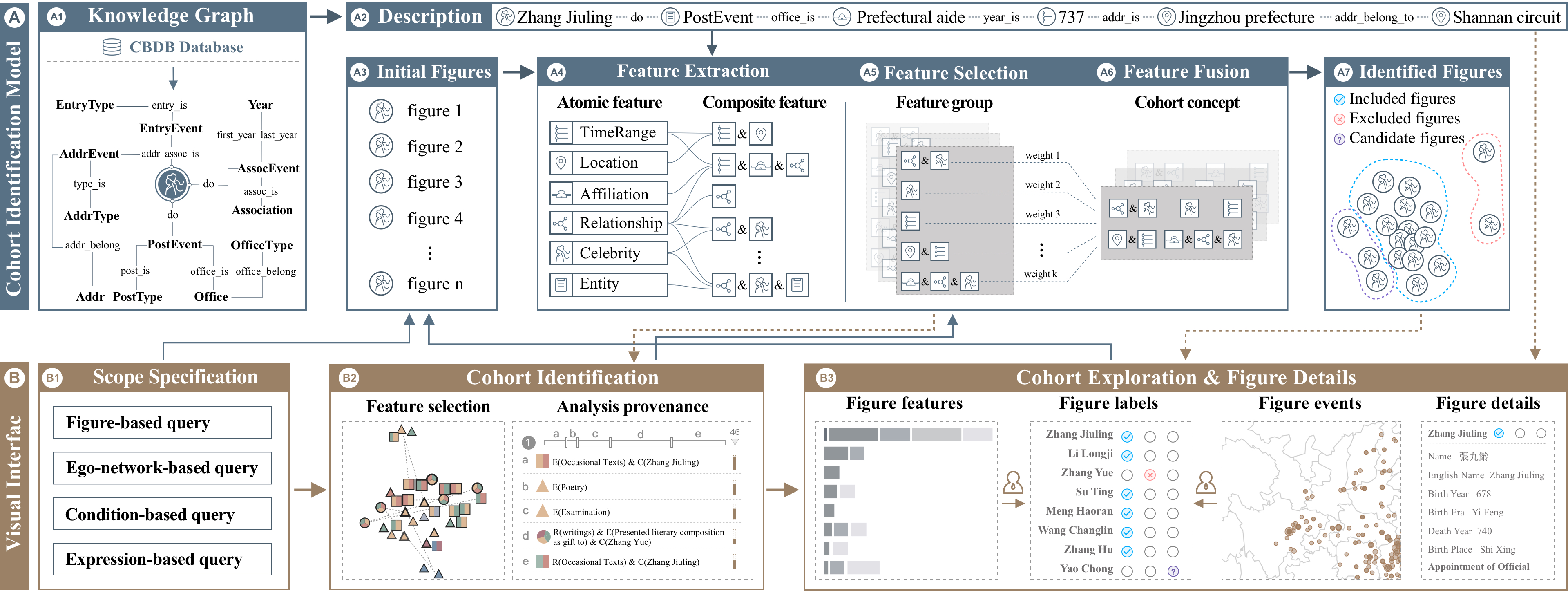}
\caption{CohortVA consists of (A) a cohort identification model and (B) a visual interface. We first build (A1) a knowledge graph based on the CBDB and extract (A2) the descriptions. Here, the example from CBDB is translated with~\cite{hucker1985dictionary} to help with the interpretation. Historians specify (A3) the initial figures from (B1) the scope specification component. The figures and extracted descriptions are piped into the model for (A4) feature extraction, (A5) feature selection, and (A6) feature fusion. 
Then our model automatically identifies (A7) the cohort figures. Generated features and figures are presented in (B2) the cohort identification component, (B3) cohort exploration and figure details component for further interpretation and exploration. }
\label{fig:datapipeline}
\vspace{-0.2in}
\end{figure*}

\rc{At the beginning of cohort explorations, users determine a search scope and specify an initial cohort (\autoref{fig:datapipeline}B1).
These initial figures (\autoref{fig:datapipeline}A1) are the query bases of cohort identification.}
We generate atomic and composite features from their descriptions. 
\rw{
For instance, given the description ``Zhang Jiuling served as the prefectural aide in 737 at the Jingzhou prefecture", we use the $TimeRange$ model (see~\autoref{tbl:atomic_feature}) to extract the time range feature $TimeRange(737)$ and the $Location$ model for location feature $Location(Jingzhou)$. 
The two atomic features jointly form the composite feature $[TimeRange(737)~\&~Location(Jingzhou)]$. 
In addition, since Zhang Jiuling is a renowned scholar-official (further descriptions in~\autoref{case2}), he is extracted as the feature $Celebrity(Zhang~Jiuling)$. 
The three atomic features then generate another composite feature $[Celebrity(Zhang~Jiuling)~\&~TimeRange(737)~\&~Location(Jingzhou)]$.}

\subsection{Feature Selection}
\label{sec:feat_select}
\rw{Historical records contain numerous descriptions for figures, resulting in a large number of extracted features. However, many features are redundant and insignificant.}
\rw{We adopted the Minimum Redundancy Maximum Relevance algorithm (mRMR)~\cite{peng2005feature} to select appropriate features, which measures statistical dependencies among features.} 

Given an initial set of $n$ extracted features $\mathbf{F} = \lbrace f_1, f_2, \dots, f_n \rbrace$, where $f_i$ refers to the $i^{th}$ feature, mRMR selects $k$ features $\mathbf{F}^* = \lbrace f^*_1, f^*_2, \dots, f^*_k \rbrace$, which are the least redundant and the most significant. It can be individually formulated as follows:
\begin{equation}
  \label{eq_min_r}
  min\; R(\mathbf{F}^*) = \frac{1}{|\mathbf{F}^*|^2}\sum_{f^*_i, f^*_j\in \mathbf{F}^* \atop f^*_i \neq f^*_j} PMI(f^*_i; f^*_j)
\end{equation}
\begin{equation}
  \label{eq_max_d}
  max\; D(\mathbf{F}^*, \mathbf{F}) = \frac{1}{|\mathbf{F}^*||\mathbf{F}|}\sum_{f^*_i\in \mathbf{F}^*, f_j \in \mathbf{F} \atop f^*_i \neq f_j} PMI(f^*_i; f_j)
\end{equation}
\rw{where Eq.~\ref{eq_min_r} aims to select independent features, and Eq.~\ref{eq_max_d} aims to select significant features. The point-wise mutual information (PMI)~\cite{church1990word} is applied to measure the redundancy between two features:}
\begin{equation}
  \label{eq_pmi}
  PMI(f_i,f_j) = log\frac{p(f_i,f_j)}{p(f_i)p(f_j)}
\end{equation}
\rw{where $p(f_i)$ refers to the probability that a figure has the feature $f_i$, and $p(f_i, f_j)$ denotes the probability that the figure has both $f_i$ and $f_j$. 
Statistically, these probabilities can be estimated by the ratio of the figures containing the features.
Thus, a higher PMI indicates that the feature pair are more dependent on each other.}

\rc{We optimize both equations by a genetic algorithm~\cite{whitley1994genetic}, which would yield multiple sub-optimal solutions. 
Each solution contains $k$ features, where $k$ is defaulted at 5 to balance the model complexity and representation capacity.
We call each solution a \textit{feature group}.
For every feature $f^*_i$ in one feature group, the two features with the highest PMIs are provided as the \textit{redundant features}.}

\subsection{Feature Fusion}
\label{feat_fuse}

The selected $k$ features allow each figure to be represented as a $k$-dim feature vector $\bm{v} = [v_1, v_2, \dots, v_k]$, where $v_i$ is the \textit{frequency} of $f_i$ and $v_i = \frac{N_{f_i}}{N_{d}}$, 
where $N_{f_i}$ is the number of the figure's descriptions containing feature $f_i$, and $N_{d}$ is the total number of the figure's descriptions. 

The \textit{concept} of a cohort is defined as the fusion of selected features (\autoref{fig:datapipeline}A6). 
It extends the feature group by assigning a fusion weight to each feature.
We propose a weakly-supervised classifier to determine whether a figure belongs to the cohort and learn the fusion weights. 
We calculate the \textit{Cohort Score (CS)} for each figure $\bm{v}$, having
\begin{equation}
  \label{eq_fusion}
  CS(\bm{v}) = \bm{w}^T \bm{v} = \sum^{k}_{i=1}w_i v_i
\end{equation}
\rw{where $\bm{w}$ is the fusion weights and $\bm{w} = [w_1, w_2, \dots, w_k]$. 
The classifier is implemented as a linear regression model for its high interpretability and the continuous outputs for ranking purposes. 
The initial figures (\autoref{fig:datapipeline}A3) are the positive training samples.
The stochastic gradient descent (SGD) optimizer is adopted in the learning process.}

\rw{A figure's membership in a cohort thus depends on its similarity to the cohort's concept.
The higher the cohort score, the more likely the figure belongs to the cohort.
To account for a possible mismatch from the extracted features and specified figures, every figure will be reassigned a new label regardless of the initial group specification \textbf{(T2)}. 
A figure with a cohort score over 1.0 is included in the cohort. 
The one whose cohort score is below 1.0 and over 0.5 is viewed as a candidate, and others are excluded.
A new cohort is then identified (\autoref{fig:datapipeline}A7) and presented to historians for visual exploration (\autoref{fig:datapipeline}B3).}

\section{Visual Analytic System}
\label{Visual}

We propose CohortVA to present the cohort identification results with explanations and to support iterative exploration. 

\subsection{Two-stage cohort analysis workflow}
\rc{
CohortVA, a visual analytics system, follows a two-stage cohort analysis workflow as shown in~\autoref{fig:workflow_comparison}B.
To establish the scope of study~(\textbf{T1}), the~\textit{\viewA~Component} (\figpipe{B1}) supports historians in specifying an initial group of figures according to their research interests. The initial group is then piped into the workflow.
}

\rc{\textbf{Cohort identification stage.}
Given a defined group, CohortVA produces a series of cohort candidates using the cohort identification model~(\textbf{T2}). 
The concepts and features of the identified cohort candidates are validated in the~\textit{Cohort Identification Component} (\figpipe{B2})~\textbf{(T3)}.
Historians can compare different concepts and adjust the features' weights in the \textit{\viewBb~View}. 
They can also replace certain features with their redundant features in the \textit{\viewBa~View}.
After validating the concepts, historians select and focus on a cohort candidate for further analysis.
}

\rc{\textbf{Cohort exploration stage.} }
\rc{CohortVA lets historians explore and refine the selected cohort from two perspectives: the cohort concept and included figures.
The figures are described by the~\textit{\viewC~Component} and the~\textit{\viewD~Component}~(\figpipe{B3}). 
Historians can cross-check the cohort composition according to historical events\textbf{~(T4)} and detailed figure descriptions\textbf{~(T5)}. 
After validation, CohortVA supports historians in excluding figures from the cohort and including related ones in the~\textit{\viewCbb~View}.
}

\rc{
Adjustments from either perspective will update the cohort interpretation for the other.
Thus, historians can iteratively refine a cohort by piping it back to the Cohort identification stage~(\textbf{T6}). The analysis process is recorded in the~\textit{\viewBb~View} for backtracking and testing what-ifs~(\textbf{T7}). When the analysis result is satisfactory, historians can export the cohort concept and the list of included figures in CSV format.
}

\subsection{\viewA~Component}
\label{ViewA}
The~\textit{\viewA~Component} (\fig{A}) provides a control panel for data queries to help historians quickly locate a target figure group. 

{\bf Enabling flexible group queries.} 
Four means of figure queries are supported: 1) figures-based query, searching figures by name; 2) ego-network-based query, starting with a core figure and expanding with related figures; 3) condition-based query, picking out figures with common descriptions such as year, location, and identity; and 4) expression-based query, allowing skilled historians to write expressions in the feature representation format (\autoref{sec:feat_extract}) to search for figures, which is most flexible but challenging.

\subsection{\viewB~Component}

The~\textit{\viewB~Component} (\fig{B}) visualizes the features of the recommended cohorts for historians to select an appropriate cohort and conduct the following exploration.
The group features of the search scope and the generated cohort identification schemes are presented in the~\textit{\viewBa~view} and the~\textit{\viewBb~view}, respectively.

\subsubsection{\viewBa~View}

The~\textit{\viewBa~View}~(\fig{B1}) interprets all identified features from the initial group as the cohort concept. To facilitate concept understanding, this view depicts each feature's importance, similarity, and overall distribution in feature categories. We encode the features in different channels to make a distinction. 

\begin{figure}[t]
\centering
\includegraphics[width=\linewidth]{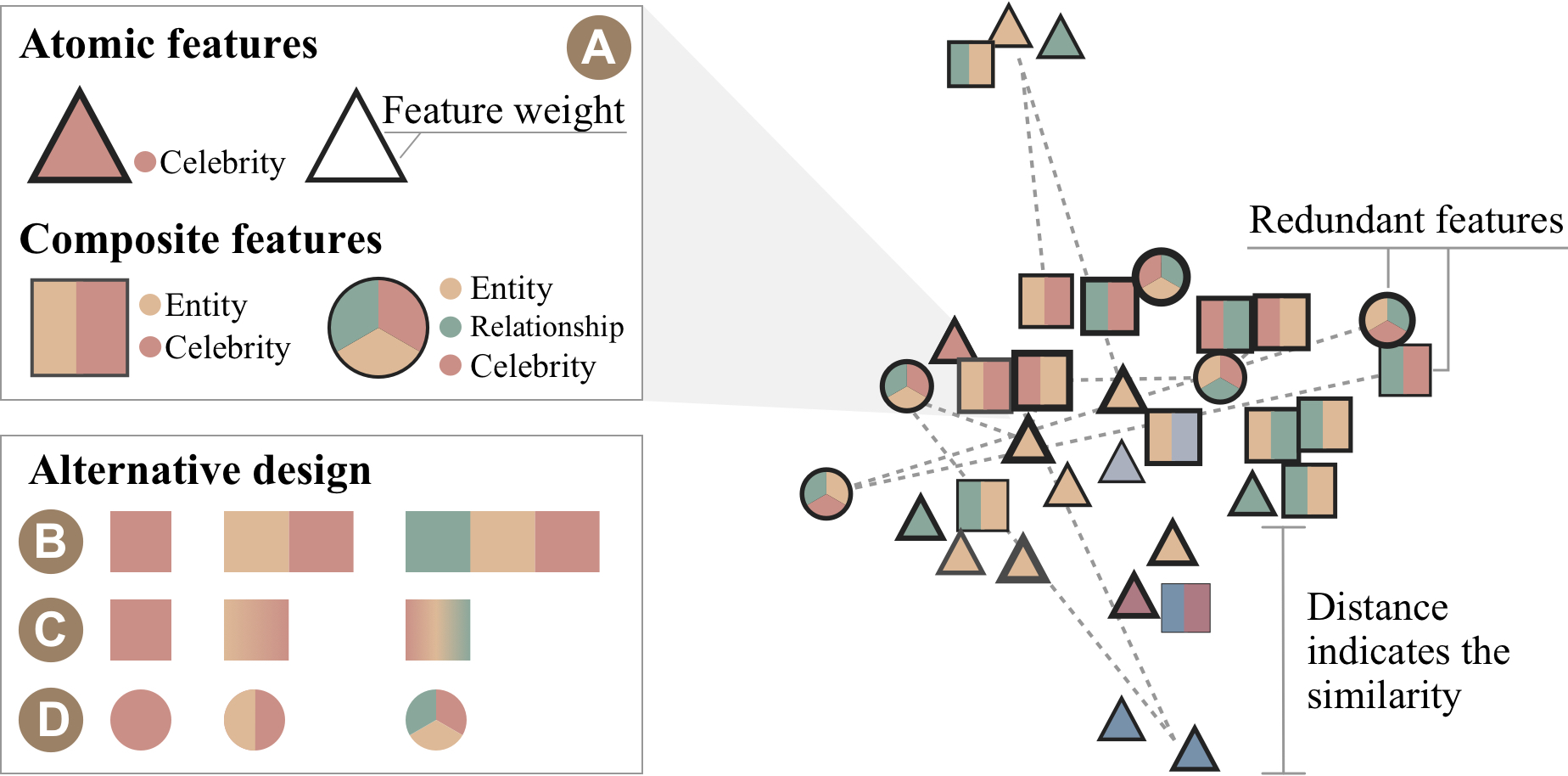}
\caption{Cohort Feature Selection View. (A) The encoding scheme for atomic and composite features. (B, C, D) The alternative designs.}
\label{fig:alter_design}

\end{figure}

{\bf Visualizing features in colors, shapes, and thickness.} 
The extracted features could be either atomic or composite (see \autoref{sec:feat_extract}).
We use different colors to distinguish the six types of atomic features (\ie, \feature{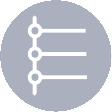}{TimeRange}, \feature{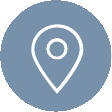}{Location}, \feature{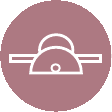}{Affiliation}, \feature{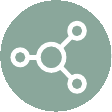}{Relationship}, \feature{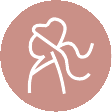}{Celebrity}, and \feature{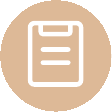}{Entity}). 
The composite features combine the colors of their corresponding atomic features. 
We explicitly encode the number of atomic features by shapes to emphasize the distinction.
Specifically, triangles represent atomic features, while squares and circles represent composite features consisting of two and three atomic features, respectively (\autoref{fig:alter_design}A). 
The border's thickness encodes the feature's weight.

\textit{Justification.}
\rc{We considered three alternative designs. The first one (\autoref{fig:alter_design}B) employs squares to represent atomic features and groups multiple squares to represent composite features. However, composite features with three atomic features occupy too much space and cause visual confusion. We also visualized composite features with gradient colors (\autoref{fig:alter_design}C) and color combinations (\autoref{fig:alter_design}D) of their atomic features.
However, the color differences are too small to be perceived and distinguishable.}
Therefore, we use two visual channels (\ie, shape and color) to enhance the distinction and perception of features.

{\bf Displaying features' similarities.} The extracted features are displayed in the force-directed layout. \rc{The distance between two features is proportionate to their reciprocal PMI value (see Eq.~\ref{eq_pmi}). Therefore, the distance between two features positively correlates with their similarity. If a feature has redundant features, we link them by dashed lines. As well as the graphical representation, a feature list is shown to view features sequentially. In the list, features can be sorted by the count of corresponding figures or the feature's significance (see Eq.~\ref{eq_max_d}).}

\subsubsection{\viewBb~View}

The~\textit{\viewBb~View}~(\fig{B2}) visualizes an overview of multiple identified cohorts and tracks analysis provenance.

\rc{{\bf Explaining the cohort concept.} The view lists features' descriptions, weights, and the number of related figures). Historians can replace a feature by selecting a redundant feature and clicking the ``replace" button in the \textit{\viewBa~View}.
Each feature's fusion weight can be set via the slider on top of the view. }

\rc{{\bf Recording exploring iterations.} Iterative cohort analysis processes may yield multiple versions of the cohort. For each version, the view summarizes the number of total figures, changed figures, and cohort features for cohort comparisons across iterations.}

\begin{figure*}[ht]
\centering
\includegraphics[width=1.0\linewidth]{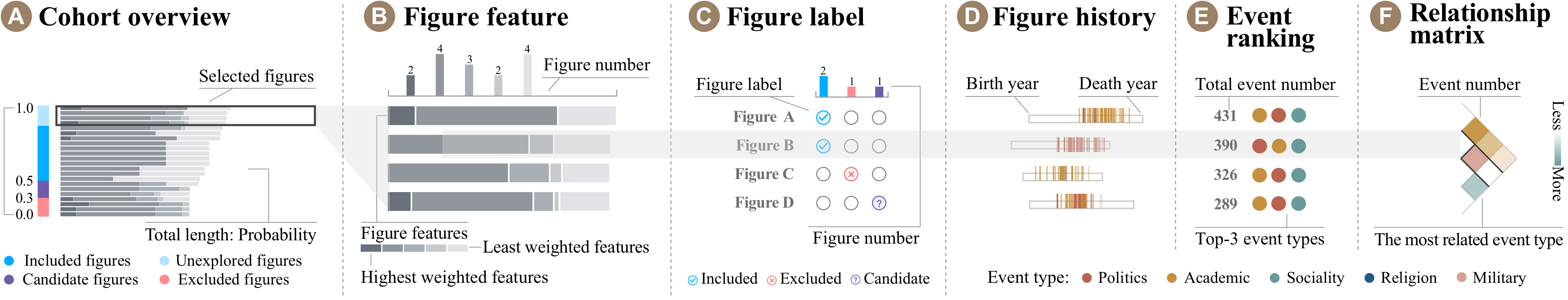}
\caption{The Cohort Exploration component. Historians can shortlist figures from (A) the cohort overview. The (B) figure features, (D) figure history, (E) figure events, and (F) figure relationships views provide supporting information for feature- and event-based validation. After cross-checking from both perspectives, historians can label the figures in the (C) figure label view.}
\label{fig:event_design}
\vspace{-0.2in}
\end{figure*}

\subsection{\viewC~Component}
\label{sec:explanation}

\rc{To enhance the interpretability of analysis results, the~\textit{\viewC~Component}~(\fig{C}) supports historians in validating cohorts from the model and data perspectives.}

\subsubsection{\viewCa}
The \textit{Cohort Overview}~(\figevent{A}) shows the cohort score (see Eq.~\ref{eq_fusion}) and the labeling status of all figures in the selected cohort. Considering data scalability issues, historians need to select a sub-group of figures of interest and check their descriptions in the~\textit{\viewCb~View} and the~\textit{\viewCc~View}.

\rc{{\bf Summarizing the feature distributions.} The view shows the feature distribution of each figure in the selected cohort. The cohort features confirmed by historians in the~\textit{\viewBb View} are distinguished in this view by different grayscale values. For a figure, the length of each feature represents the frequency multiplied by the corresponding fusion weight (\autoref{feat_fuse}). To focus on a certain feature, historians can sort figures according to the feature's value.}

\rc{{\bf Summarizing figure labels.} CohortVA employs a multicolored bar to group figures included in the cohort (colored in \feature{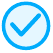}{blue}), candidate figures (colored in \feature{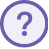}{purple}), and excluded figures (colored in \feature{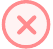}{pink}) on the left of the view. It shows the status of the labeling progress and gives an overview of the labeled figures.}

\textit{Justification.} 
\rc{We tried to encode the figure distribution with colors used in the~\textit{\viewBa}~(\fig{B1}). However, the multiple colors of composite features did not scale well with the large number of figures, similar to the alternative design in \autoref{fig:alter_design}B. The numerous colors distracted historians from other views, and the limited screen space also poses challenges in distinguishing features. 
Thus, we unified the features and colored them with different shades of gray.}

\subsubsection{\viewCb~View}

\rc{The \textit{\viewCb~View}~(\fig{C2}) shows model-related information of the sub-group selected in the~\textit{\viewCa View} in a dual column structure. The \textit{\viewCba~View}~(\figevent{B}) in
the left column illustrates a detailed feature distribution. Besides the zoomed-in figure distribution, a histogram is employed to reflect the number of figures with each feature. The \textit{\viewCbb~View}~(\figevent{C}) in the right column lists the labels of each figure. Historians can modify labels and guide the cohort identification model to update the cohort.}

\subsubsection{\viewCc~View}
\rc{The \textit{\viewCc~View}~(\fig{C3}) visualizes the life experience of each figure by demonstrating five categories of events (\ie., politics, academic, religion, sociality, and military) of concern for historians. We use color encoding to distinguish the five categories. Detailed event descriptions from five perspectives are described below:}

\begin{itemize}[itemsep=2pt,topsep=0pt,parsep=0pt,leftmargin=*]
\item \textit{Category:} \rw{Each row in the \textit{\viewCca~View}~(\figevent{D}) shows the events in a historian-selected category of a figure. The events are visualized by a thin bar, of which the horizontal position encodes the time of the event. A row with dense bars indicates that the corresponding figure was recorded significantly in the category. The time spans of all rows are aligned to support event comparisons.
}

\item \textit{Frequency:} The \textit{\viewCcb~View}~(\figevent{E}) shows the total number of events and the top three categories of events ranked by quantity. It provides an overview of the figure's identity characteristics.

\item \textit{Relationships:} The \textit{\viewCcc~View}~(\figevent{F}) employs a 45-degree-rotated matrix to show the relationship among figures. The element in the $i^{th}$ row and $j^{th}$ column represents the events involving the $i^{th}$ and $j^{th}$ figure. The shades of color encode the event quantity. 
\rc{Since the ordering of matrix visualization has an extensive influence on local structure discoveries~\cite{nathan2022matrix}, we adopted the Girvan Newman algorithm~\cite{girvan2002community}, a betweenness-based community detection method, to sort the grids and highlight figures with closer relationships.}

\item \textit{Location:} The \textit{\viewCcd~View}~(\fig{C3-4}) shows the geographic distribution of all events by circles. The circle size encodes the number of events that happen at a location. Clustered circles highlight the region's importance to the cohort.

\item \textit{Time:} The \textit{\viewCce~View}~(\fig{C3-5}) shows the temporal distribution of events. If historians choose a specific year, the detailed information of events happening this year will be displayed for further exploration and validation.

\end{itemize}

\subsection{\viewD~Component}

The \textit{\viewD~Component}~(\fig{D}) provides historians with detailed descriptions (\ie, background information and hyperlinks to original source) of a selected figure. 
The figure's identified features are listed in the view for reference. 
The information helps historians develop an in-depth understanding to make a labeling decision.
\section{Evaluation}
\label{Evaluation}
We conducted two case studies and eight expert interviews to verify the effectiveness and usefulness of our CohortVA.

\subsection{Case Studies}
\label{case1}
We invited the historians mentioned in~\autoref{Task} to explore CohortVA freely according to their research interests and intentions. \rc{We encouraged historians to adopt the think-aloud protocol and recorded how they used our system, as described in the following two cases.}

\subsubsection{Case1: Verify the Neo-Confucianism in Song}
The research interest of H4 lies primarily in Neo-Confucianism theory in the Song dynasty. To verify the cohort identified by the traditional prosopography workflow, H4 leveraged CohortVA to explore the \rw{Neo-Confucian} cohort in the Song dynasty.

\textbf{Specify the initial cohort of interests.} 
With clear goals in mind, H4 first initialized the figure group through the conditional queries of `Song' and `Neo-Confucianists' in the \textit{\viewA~Component} (\textbf{T1}). \rc{CohortVA returned 588 figures satisfying the conditions}. Then, the cohort identification model (\autoref{sec:model}) recommended cohorts in the \textit{\viewB~Component} (\fig{B}) for further analysis (\textbf{T2}).

\textbf{Cohort identification.} To select an appropriate cohort candidate, H4 first observed the cohort feature distribution in the \textit{\viewBa~View} (\figsub{B}{1}), where several feature clusters appeared. These clusters contained atomic features (\eg, \feature{Entity}{Writings}, \feature{Entity}{Neo-Confucian}, \feature{Celebrity}{Zhu Xi}) and composite features that are mostly related to \feature{Celebrity}{Zhu Xi}, the most famous Neo-Confucianist in the Song dynasty (\textbf{T2}). 
H4 also noticed an outlier feature \feature{Location}{Fujian Lu}. 
H4 indicated that this location feature is significant to the Neo-Confucian cohort because `Zhu Xi' was born and raised in `Fujian Lu.'
Moreover, after `Zhu Xi' resigned from the government, he ran colleges to preach Neo-Confucianism in `Fujian Lu' for forty years.
Since `Zhu Xi' was the core figure, H4 further explored the cohort candidate related to \feature{Celebrity}{Zhu Xi} in \figsub{B}{2}. One of the five identified features of the cohort candidate is \feature{Location}{Song Dynasty}, which is too coarse to filter effectively. Thus, H4 replaced it with the redundant feature \feature{Location}{Fujian Lu} (\textbf{T3}).

\textbf{Cohort exploration.}
After determining the cohort features, H4 started to validate the features and refine the figures in the \textit{\viewC~Component} (\fig{C}).
In the \textit{\viewCa} (\fig{C1}), H4 re-sorted the figures according to the feature \feature{Entity}{Writings}, because authorship was an important characteristic in Neo-Confucianists. \rc{Then, H4 screened out the 226 figures with the least cohort scores,} and clicked the `academic' button to observe the writing events that occurred in their lives in the \rw{\textit{\viewCca~View}} (\figsub{C}{3-1}). It turned out that these figures rarely authored writings or participated in academic events, so H4 excluded them from the cohort \textbf{(T4)}.
    
As shown in the \rw{\textit{\viewCbb~View}} (\figsub{C}{2-2}), the figures with the highest cohort scores include several representative Neo-Confucianists, such as `Zhu Xi,' `Lu Jiuyuan,' `Lv Zuqian,' and `Huang Gan.' 
However, H4 spotted a few included figures that are not Neo-Confucianists when browsing figure descriptions.
For instance, the figure `Xin Qiji' has not contributed to the spread of Neo-Confucianism theory. `Xin Qiji' was misidentified due to his close political relationship with `Zhu Xi,' as shown in the \textit{\viewCcc~View} (\fig{C3-3}). H4 manually excluded the figure after validating with the events (\textbf{T5}). To focus on core Neo-Confucianists and reduce cohort size, H4 also excluded 337 less significant figures with less than one hundred events.

Next, H4 checked the spatio-temporal descriptions of the figures.
The \textit{\viewCcd~View} (\figsub{C}{3-4}) demonstrated that figures in this cohort had recorded events in more than 500 places. 
Most of the events were clustered in `Fujian Lu,' which proves the importance of the feature \feature{Location}{Fujian Lu}.
In the \textit{\viewCce~View} (\figsub{C}{3-5}), the time feature \feature{TimeRange}{1177,1181} attracted H4's attention due to a sudden surge in events. 
After hovering over the bar corresponding to the time range, H4 found that all these events involve `Zhu Xi.' 
For example, `Zhu Xi' re-built the Bailudong College~\cite{weihan2009analyse} (one of the four ancient academies in China) in 1179. H4 considered it a milestone event for the spread of Neo-Confucianism.

\begin{figure}[htbp]
\centering
\includegraphics[width=\linewidth]{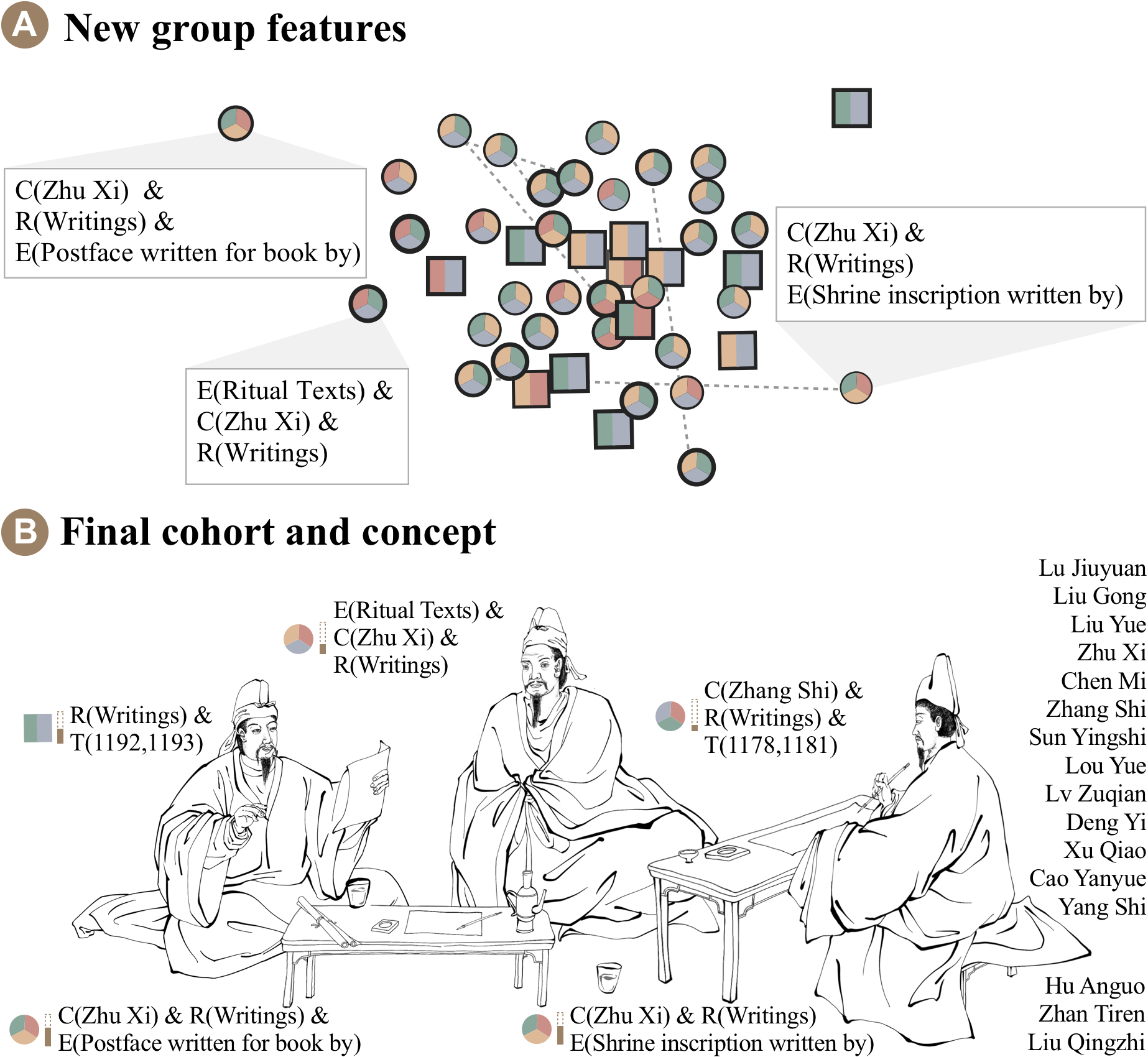}
\caption{The identified cohort and concept for Case1. (A) An overview of the new features after the first update. (B) The final cohort and concept.}
\label{fig:case1}

\end{figure}

\textbf{Iterative exploration.}
Through the above steps, H4 clicked on the `update' button in the \textit{\viewC~View}~(\figsub{C}), then the system performed the second automatic identification and recommendation in the \textit{\viewBb~View}~(\figsub{B}{2}) (\textbf{T6}). 
New features were selected based on this new cohort.
In the \textit{\viewBa~View}~(\figsub{B}{1}), H4 found more interesting features, such as \feature{Entity}{Ritual Texts} (the literature embodying Neo-Confucian theories), \feature{Entity}{Postface written for book by}, and a more precise time feature~\feature{TimeRange}{1178, 1181} (\autoref{fig:case1}A). H4 added them into the cohort features (\textbf{T6}). H4 checked the recommended figures in the \textit{\viewC~View} and verified that they align with the cohort concept.

Lastly, H4 was satisfied with the cohort having 23 core figures and the more precise concept, as shown in~\autoref{fig:case1}B.
Together with some interesting features in previous iterations (\textbf{T7}), they were exported in CSV format from the \textit{\viewBb~View}~(\figsub{B}{2}). 
\rc{H4 would further investigate how these figures spread Neo-Confucianism in the Fujian Province. 
The exported cohort would be cross-checked with other data sources (\eg, local chronicles in \textit{the Intelligent Antiquities Platform}~\cite{csab}) and verified with other research literature.}

This case shows that CohortVA can help historians quickly identify the central figures and important events in the cohort. It provides historians with new perspectives to study the spread of theories.

\begin{figure*}[htbp]
\centering
\includegraphics[width=1.0\linewidth]{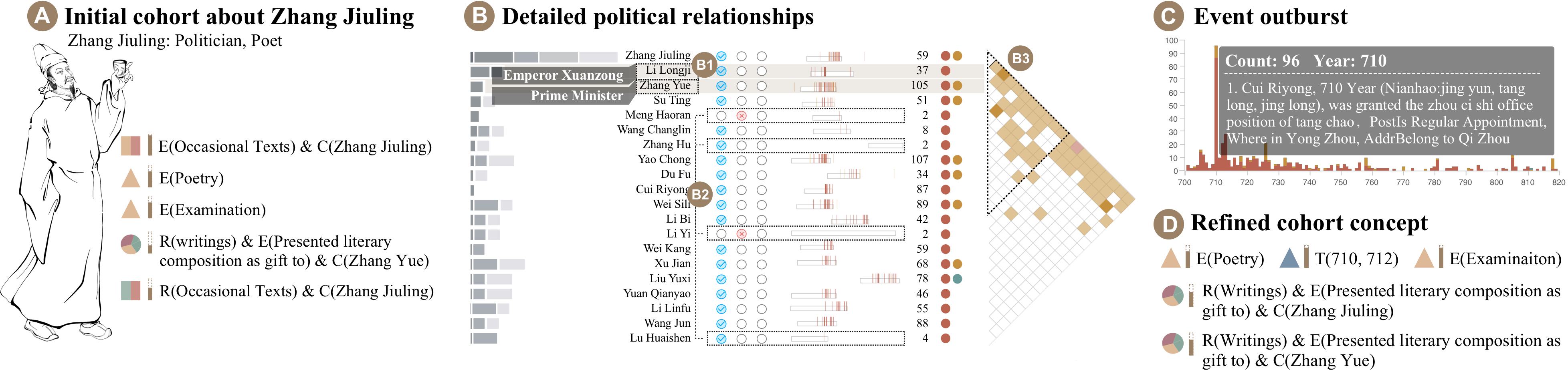}
\caption{Exploration process. The initial figures and features of Zhang Jiuling (A) were identified. Then H2 explored the political and academical relationships between these figures (B), and noticed an event outburst (C). After an update, H2 obtained the refined cohort concept (D).}
\label{fig:case2}
\vspace{-0.2in}
\end{figure*}

\subsubsection{Case2: Explore the Politics in Poetry}
\label{case2}

H2 is interested in the history of the Tang Dynasty.
In this case, H2 explored the associations around `Zhang Jiuling,' a famous scholar-official and poet, and the corresponding social influences.

\textbf{Specify the initial cohort of interests.} 
H2 used the figure-based query to choose figures related to `Zhang Jiuling' in the \textit{\viewA~View} (\textbf{T1}).
\rc{46 figures were selected from 500K historical figures}.
Then, in the \textit{\viewBa~View}, CohortVA recommended several cohort candidates whose features were mainly political and literary (\textbf{T2}). 
H2 selected a cohort with the most political and literary features in the \textit{\viewBb~View} for further exploration (\textbf{T3}).
Besides the feature~\feature{Celebrity}{Zhang Jiuling}, H2 noticed three identified features of this cohort: 1) \feature{Entity}{Examination} indicating that the figures had become bureaucrats through the imperial examinations; 2) \feature{Location}{Jingzhao} reflecting that this group had frequently been active in `Jingzhao Fu,' the official designation of the Tang's dynastic capital; 3) \feature{Entity}{Presented literary composition as a gift to} demonstrating the prevalent literary culture among the cohort (\autoref{fig:case2}A) (\textbf{T3}).

\textbf{Cross-checking and refining the selected cohort.}
H2 selected the figures with the highest cohort scores and checked their profiles (\textbf{T5}).
H2 found that a large proportion of these figures were high-ranking officials. For example, `Li Longji' and `Zhang Yue' (\autoref{fig:case2}B1) were the emperor and the prime minister of the Tang dynasty, respectively.
The event distributions in the \rw{\textit{\viewCca~View}} showed that the political and academic events were the most frequent event types (\textbf{T4}). Combined with the relational information in the matrix (\autoref{fig:case2}B3), H2 concluded that the tight clique of `Zhang Jiuling' had been deeply associated with academics and politics, especially the political power center of the Tang Dynasty. H2 removed the figures with \rc{less than five} academic or political events recorded (\autoref{fig:case2}B2).

After clicking the `update' button, the percentage of figures with the feature \feature{Entity}{Examination} had increased, and a new feature \feature{Entity}{Poet} had appeared (\textbf{T6}). In the \textit{\viewCcc~View} (\autoref{fig:case2}B3), most of the interactive events about these figures were related to `Poetry as a gift,' a more precise description than `literary composition.' 
From the analysis result, H2 suspected that poets could bond with officials in the Tang Dynasty by presenting poetry.
Favored by the officials, these poets could gain an advantage in the imperial examination for selecting state bureaucrats.
For example, `Zhang Jiuling' ingratiated himself with `Zhang Yue' and `Li Longji' by writing poems.
When he became the prime minister, other poets started writing poems for him as well.
Besides, a new feature~\feature{TimeRange}{710, 712} appeared after the update. Looking at the \textit{\viewCce~View} (\autoref{fig:case2}C), H2 found that most figures of this cohort gained government positions during this period, \rc{where 96 events were recorded}. H2 recalled that the Tang Long coup occurred in 710, after which the regime alternation caused significant bureaucratic alternations.
After comparing with other cohorts in previous iterations, H2 decided to include this feature \textbf{(T7)}.

\textbf{Export the exploration results.}
Finally, H2 exported the exploration results.
H2 obtained a refined cohort (consisting of 17 figures) and derived its completed concept as in \autoref{fig:case2}D.
\rc{H2 wanted to determine if this cohort was united in their political views. 
Therefore, he will consult the official documents of the Tang dynasty on \textit{The AiRuSheng Platform}~\cite{airusheng} to further explore the influence of this cohort in politics.}

This case shows that CohortVA enables historians to obtain a more comprehensive understanding of the cohort and discover hidden connections between features.

\subsection{Historian Reviews}
\label{sec:feedback}

To evaluate the effectiveness of CohortVA, we invited eight historians (\ie, three professors (H1-H3) and five PhDs (H4-H8)) to participate in the interview. 
\rc{We collaborated with H1-H5 for two years, as mentioned in~\autoref{Background}. They participated in multiple design iterations of the CohortVA. Three historians (H6-H8) work on historical research on the Tang, Song, and Ming dynasties, respectively, using CohortVA for the first time.}
H1-H6 are familiar with the CBDB and digital tools (\eg, Gephi and Netdraw) for historical research. 
H7 and H8 indicated that they mainly employed paper-based historical documents in their daily research. We interviewed three historians (H1-H3) online. Face-to-face interviews were conducted for others (H4-H8). None of the eight participants were co-authors of this manuscript.

\textbf{Procedure.} 
Initially, each interviewee was asked to fill out a consent form and a demographic questionnaire about their background. Then, we completed the following steps to collect their comments.

\begin{itemize}[itemsep=2pt,topsep=0pt,parsep=0pt,leftmargin=*]
    \item{\emph{Training (20 min)}. We introduced our motivation, related definitions, the cohort identification model, and visual designs following the visualization introductory description guidelines~\cite{yang2021explain}.}
    \item{\emph{Freeform Exploration (45min)}. We let historians explore our system freely. During the exploration, historians were encouraged to adopt the think-aloud protocol. Their exploration processes were recorded.}
    \item{\emph{Interview (20 min)}. We asked them to evaluate our system from three aspects, \ie, the effectiveness of the approach, the reliability of the result, and the usability of the system.}
\end{itemize}

\textbf{Effectiveness of the approach.} 
All historians agreed that their research could benefit from the proposed analysis approach.
Five historians (H2, H4-H7) pointed out that CohortVA can significantly improve research efficiency.
H5 mentioned that they usually missed significant features and made tremendous efforts to re-identify features after initial explorations. Automatic feature extraction can free them from exhausting concept extractions and allow them to focus on other important tasks, like analyzing the causes and effects.
H4 said, ``CohortVA acts like a reference book that I will refer to at different research stages.''
Despite the preference for traditional research methodologies, H8 was interested in using CohortVA to explore unfamiliar areas for efficient explorations and inspiration.
H3 affirmed the significance of our work and believed that CohortVA is an appropriate teaching tool for history classes.
Besides, H2 suggested that our approach should further support comparing multiple cohorts, which would be one of our future works.

\textbf{Reliability of the result.}
Historians must validate the model output before applying the output in historical research. 
H5 and H7 indicated that the \textit{\viewC~Component} (\fig{C}) favored them in understanding the extracted cohort features.
As mentioned in \autoref{case1}, contextual information helps historians understand why \feature{Location}{Fujian Lu} is included in cohort features.
H5 and H6 argued that visual designs and interactions facilitate the connection between interpretation and context. 
\rc{H2 and H4 indicated that adding a hyperlink in the \textit{\viewD~Component} (\fig{D}) was a great function for checking the original material.}
\rc{H3-H8 pointed out the limitation of a single data source and expressed the wish to introduce more data sources, including user-defined data for cross-checking.}

\textbf{Usability of the system.} 
Historians agreed that each view was well-designed.
Among all visual forms, historians (H2-H5) considered \textit{\viewCcd~View} (\fig{C3-4}) to be the easiest to use, believing it has "the most distinct visualization characteristic and the most concise data presentation."  
H3 suggested providing maps of different historical dynasties to support exploring various research targets.
The \textit{\viewCce~View} (\fig{C3-5}), visualizing events number in each year, was also popular among historians.
Although other visualizations, such as the \textit{\viewBa~View} (\fig{B1}) and the \textit{\viewCcc~View} (\fig{C3-3}), require some learning costs for historians unfamiliar with data visualization and analysis,
H2 commented on them as being valuable visualization projects, ``digital humanities will inevitably be more complex but powerful to support complicated analytical tasks.''
H4 particularly preferred swift interaction designs, like the filtering and sorting functions in the \textit{\viewCa~View} and the tagging function in the \textit{\viewCcc~View}.



\section{Discussion}
This section summarizes the lessons learned from working with historians, as well as the limitations and future work of CohortVA.
Through collaborations with historians, we have summarized three design implications for digital humanities (\textbf{CAD}): 

\textbf{Comprehensibility.} 
Historians prefer easy-to-understand and simple visualizations. Being used to reading a large literature base, they feel more familiar with concrete wordings than abstract visual forms. 
\rc{In our design, we contextualize the generated descriptions with conventional visualizations to simplify the paradigm shifts from reading to perceiving.
Moreover, historians are more willing to adopt visualizations that require higher learning costs when presented with clear usage benefits. 
For example, for the \textit{Relationship Matrix View}, we used a matrix ordering algorithm to cluster similar groups and encoded relationship types with colors. 
The efficiency in pattern mining has drawn historians' interest in studying.
As H3 indicates, ``we can quickly discover the types of social relations between figures through the colors on the matrix, which is very novel." 
However, visual uncertainty should be used with caution because some historians are conservative about probabilistic inference and favor definite evidence.}


\textbf{Authenticity.} 
We have identified four ways to enhance historians’ trust and confidence in the method, process, result, and data.
\rc{
1) Reproducing cohorts that \textit{agree with existing theories} can increase the methods' trustworthiness. It directly verifies functional completeness and correctness.
2) Steering results iteratively give historians more confidence in the process. In CohortVA, historians gradually apply their \textit{prior knowledge and expertise} to annotate and validate the recommended figures. The smaller gaps between iterations help them infer the different results.
H3 highly appreciates the provenance tracking function, ``which helps me to trace back and adjust my hypotheses for further research."
3) Providing organized information across various perspectives enhances the result's confidence. Historians emphasize \textit{cross-checking} the automatic analysis results to reduce biases and misinformation.
4) Displaying and providing access (\eg, hyperlinks) to the \textit{original data sources} are important features in building trust in the data. Historians value the authenticity of the original textual documents.}

\textbf{Diversity.}
\rc{
Since the ancient documents and historical data span thousands of years, special attention should be paid to the diversity in the underlying spatial and temporal contexts. 
For spatial contexts, map visualizations should be aware of landscape changes.
For example, the Yellow River has undergone six major avulsions (\ie, changes in river's course) in history.
Ancient times' regional landscapes and geographical characteristics have notable differences from their modern counterparts.
Also, the territorial changes in different dynasties would create trouble recognizing and understanding historical events with modern maps.
Therefore, we could provide \textit{terrain or historical maps} for reference~\cite{mapgis}, as suggested by H2.}

\rc{
For temporal contexts, besides the missing data and uncertainty issues~\cite{9695348}, we came across the subtle semantic changes in interpreting entities.
For instance, the ``Prefectural aide" in~\figpipe{A2} was represented as ``Zhang Shi" in CBDB. 
From the Qin to Song dynasty, ``Zhang Shi" meant the government aide, while it became the administrator from the Yuan dynasty and on~\cite{hucker1985dictionary}.
We used composite features to capture these differences implicitly and relied on domain knowledge to spot the difference.
H1 suggested that additional knowledge representations could be adopted from \textit{existing theories and structured dictionaries}~\cite{hucker1985dictionary} to provide appropriate contexts and reduce ambiguity.}


\subsection{Limitations and Future Work}
We outline the current limitations to be addressed in future work.

\textbf{Data source.} \rw{The current system only employs a single data source. Validation with multiple data sources can further strengthen the interpretability of cohort features. In the future, it should combine with other large-scale databases or self-defined knowledge bases. We could define more description templates for knowledge graph completion, but entity alignment is still a challenging problem.}

\rc{\textbf{Cohort comparison.} The current system does not provide comprehensive comparisons among cohorts. Analyzing a single cohort limits the research scope and could be biased. Cohort comparison is a complex analysis task we will pursue in our future work.}

\rc{\textbf{Quantitative evaluation. }Historians agree that our work can improve their trust in machine learning results, but the claim can benefit from a quantitative evaluation. For example, measuring the decision time and compliance with the recommended results provide viable metrics for comparison with other systems.}


\textbf{Generalizability}. 
In this work, we mainly cooperated with historians and developed a bespoke tool for them. 
Due to different requirements, the interface and specific parametric settings cannot be easily applied to other domains.
However, the CohortVA workflow is generalizable to domain-specific tasks targeting closely related groups.
By properly defining the features, it can process large text corpora and relational databases for many domains, such as medicine~\cite{vbridge}, finance~\cite{taxthemis}, and literature analysis~\cite{songci}. 
For example, social network relationships and social media posts can replace the history corpus in our system for identifying different social groups among users.

\section{Conclusion}


In this paper, we present CohortVA, an interactive visual analytic system for historians to identify and explore historical cohorts.
Given an initial group of figures, the cohort identification model in CohortVA automatically extracts their common features and identifies potential cohorts to improve the efficiency of historians' research. 
The visual interface of CohortVA provides various supporting information to help historians cross-check these results, fostering trust in the system and a deeper understanding of cohorts. Two case studies and the historian interviews demonstrate the usefulness and effectiveness of our system. 
We summarized the lessons learned for developing CohortVA and believe that these design implications will guide system designers in dealing with historical data and working with historians.

\acknowledgments{
The authors wish to thank the editors at CBDB for their feedback and contribution to this project. This work was supported by the National Natural Science Foundation of China (No. 62132017, 61972122), Shanghai Municipal Science and Technology General Program (No. 21ZR1403300) and Sailing Program (No. 21YF1402900).
}



\bibliographystyle{src/abbrv-doi}

\bibliography{main}
\end{document}